\documentclass[twocolumn,prb,amsmath,amssymb]{revtex4}
\usepackage{graphicx} \usepackage{subfigure} \newcommand{\etal}{{\it
et al.}}

\begin{document}

\title{The effect of impurities on spin polarized Zeeman bound states
in  dilute magnetic semiconductor-superconductor hybrids}

\author{Shi-Hsin Lin} \affiliation{ Department of Physics, University
of Notre Dame, Notre Dame, IN 46556, USA} \author{Tatiana
G. Rappoport} \affiliation{Instituto de F\'{\i}sica, Universidade
Federal do Rio de Janeiro, Caixa Postal 68.528-970, Rio de Janeiro,
Brazil}

\author{Mona Berciu} \affiliation{Department of Physics and Astronomy,
University of British Columbia, Vancouver, BC V6T 1Z1, Canada}

\author{Boldizs\'{a}r Jank\'{o}} \affiliation{Department of Physics,
University of Notre Dame, Notre Dame, Indiana 46556}
\affiliation{Materials Science Division, Argonne National Laboratory,
Argonne, Illinois 60439}


\begin{abstract}

We investigate the effect of single and multiple impurities on the
Zeeman-localized, spin polarized bound states in dilute magnetic
semiconductor hybrid system. Such bound states appear whenever a
dilute magnetic semiconductor showing 
giant Zeeman effect is exposed to an external magnetic field showing
nanoscale inhomogeneity.  We consider the specific example of a
superconductor-dilute magnetic semiconductor hybrid, calculate the
energy spectrum and the wave functions of the bound states in the
presence of a single impurity, and monitor the evolution of the bound
state as a function of the impurity strength and impurity location
with respect to the center of the Zeeman trapping potential.  
Our results have important experimental implications as they predict robust spin textures even
for than than ideal samples.
We find
that for all realistic impurity strengths the Zeeman bound state
survives the presence of the impurity.  We also investigate the effect
of a large number of impurities and perform ensemble averages with
respect to the impurity locations. We find
that the spin polarized Zeeman bound states are very robust, and they
remain bound to the external field inhomogeneity throughout the
experimentally relevant region of impurity concentration and
scattering strength.

\end{abstract}
\pacs{75.50.Pp,74.25.Qt,73.43.-f, 73.21.-b}
\maketitle

\section{Introduction}
Due to its potential use in spintronics and quantum information
 applications, the controlled manipulation of individual spins has
 been a subject of great interest in recent years.  In the search for
 appropriate materials for device fabrication, one possible route is
 the use of III-V or II-VI diluted magnetic semiconductors (DMS).
 
In II-VI DMSs, the exchange interaction between band charge carriers
 and the localized magnetic spins of the magnetic dopant (such as Mn)
 gives rise to a giant splitting between band states with different
 spin components. This giant Zeeman spin splitting has been
 extensively investigated and characterized.  \cite{Furdyna1} Given
 the linear dependence of this splitting on the external applied
 magnetic field for fields lower than roughly 1T, it can be described
 in terms of a very large effective \mbox{$g$-factor} for the band
 carriers. For example, Dietl \etal{}\cite{Dietl1} reported an 
 electron effective \mbox{$g$-factor} of about 500 at sub-Kelvin
 temperatures in a \mbox{CdMnSe} sample, which implies a value of
 about 2000 for the effective \mbox{$g$-factor} of a hole in this
 material. In previous works \cite{03prl,05apl,05prb,05nature,06prb}
 we proposed to take advantage of this large effective
 \mbox{$g$-factor} by combining it with a spatially inhomogeneous
 applied magnetic field, to create a spatially varying Zeeman
 potential that acts as a confining potential for only one spin
 orientation. In suitable conditions, this leads to a single charge
 carrier with a well defined spin being trapped in the region of large
 local magnetic field. This spin-polarized charged carrier can then be
 manipulated through external control on the applied inhomogeneous
 magnetic field.

The needed nanoscale inhomogeneous magnetic field can be generated in
various ways. One possibility is to use nanoscale magnets placed above
a DMS quantum well (QW).\cite{03prl,05apl,05prb,peeters,freire}
Depending on the shape and orientation of the nanomagnet, different
nonuniform fields are generated, giving rise to various types of
confined states.\cite{05prb}

Another possibility is the use of Abrikosov vortices that appear in
type-II superconductors (SCs). Above the lower critical field
B$_{c_1}$ vortices populate the superconductor, forming a vortex
lattice. Nano-engineering can be used to insure that the vortices
nucleate in well defined positions.\cite{nanosup} The field of a
single vortex is non-uniformly distributed around a core of radius
$r\sim \xi$ ($\xi$ is the SC coherence length) and decays away from
its maximum value at the vortex center over a length scale $\lambda$
($\lambda$ is the penetration depth). If a SC layer that hosts such
vortices is placed above a DMS layer (QW), the inhomogeneous magnetic
field of the SC vortices creates an inhomogeneous magnetic field in
the DMS layer.  According to our previous
calculations,\cite{05nature,06prb} these fields are sufficiently large
to result in the confinement of band carriers with a given spin
orientation in the small region of the DMS QW that is located directly
under a SC vortex core. Thus, spin textures are formed and can be
manipulated by moving the source of the magnetic field, {\em i.e.} the
SC vortex. In other words, the SC vortices act as spin and charge
tweezers and can be used for a wide array of applications, from
investigation of the Hofstadter butterfly and making spin
shuttles\cite{05nature} to generating anions of interest for
topological quantum computing.\cite{marcel}
 
In our previous work we assumed that the DMS QW is perfectly clean and
with Mn impurities distributed in a perfectly homogeneous fashion (for
instance by digital doping\cite{digitalDMS}), so
that the Zeeman potential is a smooth function mirroring the applied
magnetic field. For such systems we obtain typical binding energies of
the spin-polarized charge carrier on the order of a few
meV.\cite{05nature,06prb} 
As we show in this paper, such perfect homogeneity is not necessary.  
This is a very important result from experimental detection and application point of view, since
it opens up the possibility of investigating Zeeman localization effects
in samples grown during standard molecular beam epitaxy (MBE) conditions.
Since
the effective \mbox{$g$-factor} is proportional to the local
concentration of Mn, inhomogeneities in their distribution will result in a
spatially-varying $g(\vec{r})$ which will modify the trapping
potential. Even more detrimental are charged impurities which exist in
any sample. Charge dopants are needed to introduce charge carriers in
the II-VI DMS, because the Mn is isovalent with the valence-II element
it replaces. The disorder due to these charged dopants can
be minimized in the usual way used for two-dimensional electron
systems, by doping at some distance away from the DMS QW and using
gates to control the concentration of charge carriers in the DMS
QW. However, small concentrations of undesired charged dopants in the
DMS QW cannot be avoided.

In this article we investigate the role played by charged scattering
centers on the confinement of spin polarized carriers in the DMS
QW. We focus on repulsive centers, which are expected to be most
detrimental to the binding of the spin-polarized carrier in the Zeeman
trap, and then briefly discuss the other types of scattering centers
mentioned above. We show that due to the difference between the length
scale of the confining potential and the repulsive scattering
potentials, the bound state energy is not significantly changed even
in the presence of a large number of impurities. This result indicates
that the binding of spin-polarized charge carriers in Zeeman traps
created in a DMS QW is very robust against such defects, with
immediate implications for spintronics applications based on this
scenario.

It is important to also emphasize that our conclusions apply to a much
 wider range of systems than the one of interest to us, because these
 conclusions are based on quite simple physics, as we show below. 
 We argue that in any system (such as quantum dots,
 quantum wells, etc.) where localization occurs on a lenghtscale much
 larger than that of typical scattering potentials, the effect of such
 scattering centers is very small even if their potential is very
 large.

This paper is organized as follows: In the next section we describe in
more detail the system we study, our theoretical approach and the
approximations we have made. In order to build intuition and
to help interpret later calculations, in section \ref{results1d} we
analyze results for a simplified one dimensional (1D) model in the
presence of a single impurity. This allows us both to gauge our
computational scheme against exact solutions, and to gain some
intuition about the effect of the scattering centers. In section
\ref{results2d} we repeat the analysis for a single scattering center
for the two-dimensional system of interest to us, and discuss the
influence of the symmetries in our results. Finally, in section
\ref{random} we present results for a random distribution of
impurities and various impurity concentrations. We conclude in section
\ref{conclusions} with a summary and discussion.
\section{ Theoretical Model \label{model}}

\begin{figure}[t] 
\includegraphics[width=0.9\columnwidth]{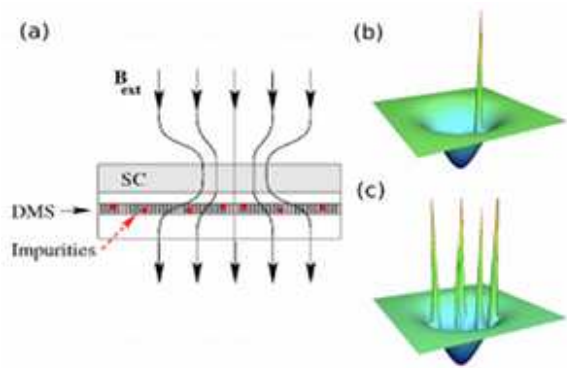}
\caption{\label{fig1} (color online) (a) Illustration of the magnetic field generated
  by a single vortex in a hybrid system made of a type II
  superconductor on top of a DMS quantum well which contains
  impurities. (b) Potential profile of the Zeeman
  trap plus  a single repulsive impurity and (c) same, but for a
  random distribution of impurities (figures not 
  to scale).}
\end{figure}

The hybrid structure of interest to us is sketched in Fig.~\ref{fig1}
(a), and consists of a SC layer in the vortex phase placed above a DMS
QW.  The two materials are separated by a thin insulating layer.  The
superconductor can be Pb or Ni, for example -- both these metals can
be grown on top of semiconductors using MBE techniques.\cite{bending}
The QW consists of a weakly p-doped DMS.  An example of such a DMS QW
is a p-doped (Cd,Mn)Te well, doped with N by a modulation doping
technique that avoids inhomogeneity effects and increases the mobility
of the carriers.\cite{dms2deg} The carriers of interest to us are
confined in the 2D DMS QW. The giant Zeeman effect in the DMS is
described by a very large $g$-factor. This combines with the
inhomogeneous magnetic field generated by each vortex to give rise to
an effective potential that binds a spin polarized hole under each
vortex, in a clean sample.
As stated, we are now interested in the effect of repulsive scattering
centers on these spin-polarized bounds states. A typical potential
well in the presence of a single, and of several charged
scatterers is illustrated in Figs. \ref{fig1}(b) and \ref{fig1}(c), respectively.

In a parabolic approximation, the Hamiltonian for a charge carrier in
the 2D DMS quantum well which interacts with the magnetic field of the
vortex and with scattering centers is
\begin{equation}
H = \frac{[\vec{p}-q\vec{A}(\vec{r})]^ 2}{2m} - \frac{1}{2}
g_{\rm eff} \; \mu_B \vec{\sigma}\cdot \vec{B}(\vec{r})+ \sum_i
V_i(\vec{r}),
\end{equation}
where $m$ and $q$ are the effective mass and charge of the carrier,
$\vec{B}(\vec{r})$ is the magnetic field generated by a single vortex,
$\vec{A}(\vec{r})$ is its vector potential and $V_i(\vec{r})$ are the
repulsive potentials from the $i$ scattering centers. For simplicity,
we assume that the motion of the carrier in the QW is two dimensional.
In the calculations shown here, we use $m = 0.5 m_e$ and
$g_{\rm eff}$=500, these being typical values for holes in II-VI
semiconductors.  Because of the large $g_{\rm eff}$, for a single
vortex the effect of the
$\vec{A}(x,y)$ term is negligible compared to that of the Zeeman
interaction and we ignore it from now on.\cite{05nature}

Experimental results show that the Zeeman splitting for holes in a DMS
is anisotropic, depending on the direction of the magnetic field with
respect to film plane.\cite{Kuhn1} Using a Luttinger
Hamiltonian,\cite{Lut} one can also show that in a 2D QW, the
$g_{\rm eff}$ of heavy holes is highly anisotropic, with an
in-plane component much smaller than that perpendicular to the film
plane.\cite{05prb} With these results in mind, we can further simplify
our problem by considering only the effect of the $z$ component of the
magnetic field, which is most strongly coupled to the charge
carrier. This approximation allows us to decouple the Hamiltonian in
the spin-up and spin-down sector, and consider them separately. We
will only focus on the spin-component for which this Zeeman potential
acts as a trap.

As discussed extensively in Ref. \onlinecite{05nature}, the size of
the confining region defined by the inhomogeneous magnetic field
depends on the superconductor's parameters and the distance between
the SC and DMS layers. It is typically several tens of nm in size. On
the other hand, the impurity potential of the scattering centers is
considerable only in a 
much smaller range of a few \AA~in the immediate vicinity of the
impurity. It is the effect of this
strong repulsive potential on the bound state that is of interest to
us, since the long-range Coulomb repulsion is too weak to be
relevant (although its effects  can also be studied with our method, if so desired).
Given the large difference between the characteristic 
length scales, we model the scattering potential as a delta function.

After all these approximations, the Schr\"odinger equation for the
trapped spin component is:
\begin{equation}
\left[\frac{{p}^ 2}{2m} - g_{\rm eff} \mu_B \frac{\sigma}{2} 
B_z(\vec{r})+ a\sum_i
\delta(\vec{r}-\vec{R}_i)\right]\psi_\sigma(\vec{r})=\varepsilon
\psi_\sigma(\vec{r}) ,\label{hamil}
\end{equation}
where $a$ characterizes the strength of the impurity potential and
$\vec{R}_i$ is the location of the $i^{th}$ impurity.

We solve for the eigenvalues in the usual way, by expanding the
wavefunction in a complete basis set and finding the appropriate
coefficients from a matrix equation. Our basis functions are B-spline
polynomials on a non-uniform knot sequence adjusted optimally for each
specific arrangement of charged impurities.  This method has been
widely used in atomic physics~\cite{bsplines} and  is ideally suited
for problems such as this. Details of the method and its advantages
are discussed in the appendix \ref{appendix}.


\section{One dimensional model \label{results1d}}

We first study a one dimensional model with a single delta
function impurity inside a square potential well. This provides an
intuitive understanding of the physics we want to explore, and also a
numerical test of our B-spline scheme since it permits the comparison
between the numerical results and the exact solution.

\begin{figure}[t]
\includegraphics[height = 1.9in]{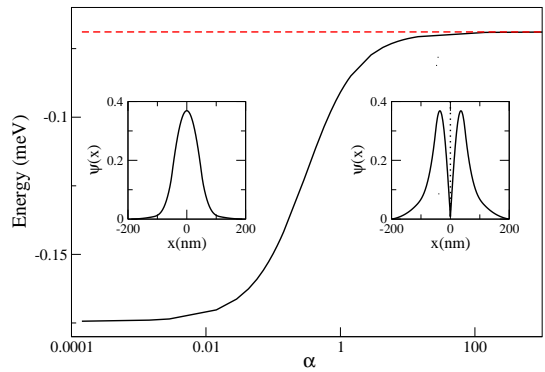}
\caption{\label{fig2} The ground state and first excited state energy
  versus $\alpha$, a dimensionless quantity related to strength of the
  impurity potential.  The impurity is located at the center of the
  well $x=0$. The left inset shows the ground-state wavefunction for $\alpha=0$, and
  the right inset shows it for $\alpha=100$.}
\end{figure}

\begin{figure}[b]
\includegraphics[height = 1.9in,clip]{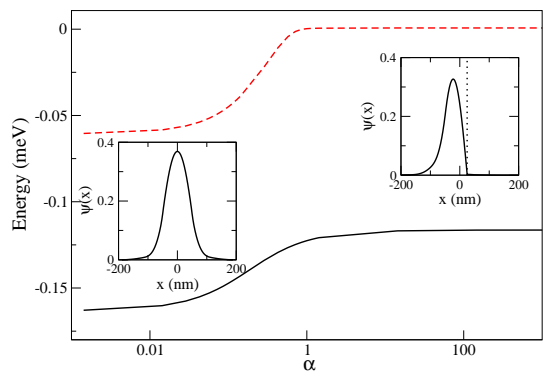}
\caption{\label{fig3} The ground state and first excited state energy
  versus the strength of the delta function impurity for $x_0=L/4$.
  $\alpha=a/a_0L$.  The left inset shows the ground-state wavefunction for $\alpha=0$, and
  the right inset shows it for $\alpha=100$.}
\end{figure}

We consider a 1D square well potential
\begin{equation}
V_0(x) = \left\{
\begin{array}{c l}
-a_0, & \mbox{if } -L/2 \le x \le L/2 \\ 0, & \mbox{otherwise}
\end{array}\right.
\end{equation}
where $L$ is the lateral size of the well. To it, we add a single
impurity potential
\begin{equation}
V_1(x) = a \delta(x-x_0)
\end{equation}
where $x_0$ is the impurity position. In this case, the
Schr{\"o}dinger equation is given by
\begin{equation}
- \frac{\hbar^2}{2m}\frac{d^2}{dx^2} \psi(x) + V_0(x) + V_1(x) =
  \varepsilon \psi(x).
\end{equation}
To understand the effects of the impurity on the bound states for
different strengths of the impurity potential, we calculated the
eigenvalues of this Hamiltonian. The energy of the lowest two bound
eigenstates is shown in Fig.~\ref{fig2} vs. the strength of the
impurity potential expressed as a dimensionless quantity $\alpha = a/
(a_0 L)$.  The parameters are $L=100$nm and $a_0=3.5 \hbar^2/(mL^2)$,
and the impurity is located in the center of the well,
$x_0=0$. Fig.~\ref{fig2} reveals that the ground-state energy first
increases with $\alpha$ and then saturates to the value of the first
excited state. On the other hand, the energy of the first excited
state is independent of $\alpha$. The reason for this behavior is
straightforward to understand. Since the first excited state has a
node at the location of the impurity, its energy and wavefunction is
naturally unaffacted by its presence. On the other hand, the original
ground state wavefunction has a maximum at the impurity location, so
 its energy increases with $\alpha$. However, for
large enough $\alpha$ the ground-state wavefunction also develops a
dip precisely at the location of the impurity and becomes insensitive
to its strength. This is demonstrated by the insets to Fig.\ref{fig3}, which show the
ground-state wavefunction at $\alpha=0$ and $\alpha=100$. In essence,
when $\alpha \gg 1$, the impurity potential splits the original well
into two isolated wells which are degenerate, given the symmetry of
the problem.

When the impurity is placed off-center, as in Fig. \ref{fig3}, the
system loses this symmetry and the $\delta$ function separates the
original well into two distinct wells of different sizes. Since the
impurity is now placed in a location where both the original ground
state and the first excited state wavefunction were finite, both
energies now increase with $\alpha$ when $\alpha < 1$. At higher
values of $\alpha$ both the ground state and excited state energies
saturate. As shown in the insets, the wavefunctions develop dips at
the location of the impurity and become progressively more localized
on one or the other smaller wells. The new ground state wavefunction
becomes confined to the larger new well.

While these results are very easy to understand, they do provide key
insight into the effects of the impurity on the bound states. First,
it is clear that symmetry plays a very important role. While this
influence is maximized by the one dimensional character of the model,
we expect analogous features in the two-dimensional case.

More importantly, we already see that even if the impurity potential
is large, its overall effect on the bound-state energy and
wavefunction is limited. Because it is extended over a much larger
area, the bound wavefunction simply avoids the impurity by developing
a dip (or a zero) in its vicinity, at relatively low energetic
cost. In 1D the impurity effectively ``cuts'' the wavefunction. This
will not happen in $d=2$ or higher dimensions. However, as we
demonstrate below, the limited effect on the binding energy is a general
feature observed in higher dimensions.

The results shown so far also illustrate why B-splines are an
ideal basis for such problems. If one tried to use a more traditional basis of orthonormal
polynomials such as harmonic oscillator eigenstates, one would have to
mix in very many basis states in order to be able to accurately
describe smooth wave-function variations on a short length scale, near the
impurity, as shown in the right inset of Fig. \ref{fig2}. Describing
regions where the wavefunction essentially vanishes, as shown in the
right inset of Fig. \ref{fig3}, would be even more difficult. On the
other hand, use of B-splines allows one to just slightly increase the number
of basis states by sampling the vicinity of the impurity on a smaller
mesh. As a result, the calculation with one or more impurities is very
comparable, in terms of numerical computational costs, with the one in
the absence of impurities. A more detailed discussion of these issues is given in
Appendix A. 


\section{Two dimensional Zeeman potential with a single impurity\label{results2d}}

We now can consider the two dimensional system in the presence of a
single charged impurity.  As we are not interested in the details of
the Zeeman potential, we take the magnetic field generated by the
vortex to have a Gaussian profile. This approximation is reasonable
close to the vortex core, where the magnetic field decays
exponentially on a length scale set by $\xi$.\cite{carneiro,05nature}
We therefore solve Eq.~\ref{hamil} where the magnetic field profile is
taken to be:
\begin{equation}
B_z(\vec{r})=B_0 \exp\left( -(\vec{r}-\vec{r}_0)^2/\xi^2\right),
\end{equation}
where $\vec{r}_0$ is the location of the vortex core, and $B_0 = 0.206
$T and $\xi = 50$ nm are the strength and the range of the magnetic
field (these are typical values for a dirty Pb film in a type II
regime).  We solve the equation numerically for a system of finite
size $200$nm $\times$ $200$nm, with periodic boundary condition and a grid
of $51\times 51$ knots.  This large lateral size is chosen so as to
eliminate finite size effects in our bound state energies.  For more
details about the implementation of this numerical method and the
periodic boundary conditions, see the appendix.

\begin{figure}[t]
\includegraphics[width= 0.8\columnwidth]{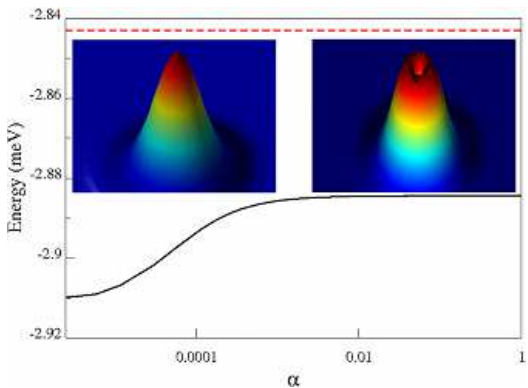}
\caption{\label{fig4} (color online) The two lowest state energies versus the impurity
  strength.  The left(right) inset is the ground-state wavefunction
  for $\alpha=0$ ( $\alpha=1$).}
\end{figure}

\begin{figure}[b]
\vspace{0.5cm} \includegraphics[height = 1.9in]{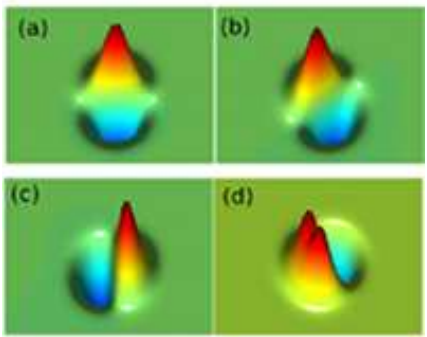}
\caption{\label{fig5} (color online) Wave-functions of the two excited states with
$p$-type symmetry when  the impurity is located at the center
of the vortex, in (a) and (c); and for an off-center impurity, in (b) and (b).}
\end{figure}

We follow a procedure similar to the one in the previous section and
begin with a single impurity located at the center of the Gaussian and
vary its potential strength. Again, we define a dimensionless quantity
$\alpha=2a/(g_{\rm eff}\mu_B B_0 \pi \xi^2)$ related to the
strength of the impurity potential. When the impurity is located at
$\vec{r}_0$, as shown in Fig. \ref{fig4} the ground state energy
increases with the $\alpha$ and saturates at a finite value for
$\alpha>0.01$.  In the insets of Fig. \ref{fig4}, we can see that the
ground-state wavefunction maintains its $s$-type symmetry as we
increase $\alpha$ but develops a dip at its center. Similarly to the
one dimensional case, the first excited state, which is two-fold
degenerate, is not affected by the impurity potential, as can be seen
in Fig. \ref{fig5} (a) and (c).  This is because its eigenfunctions
have $p$-type symmetry, with a node at the origin where the impurity
is located.  These results mirror those of the 1D case but are less
dramatic: the energy separation $\Delta$E between the two lowest
eigenstates changes by less than 30\% for arbitrarily strong repulsive
impurity potentials even when the charged impurity is placed where it
should do the most damage.

\begin{figure}[t]
\vspace{0.5cm} \includegraphics[height = 1.9in,clip]{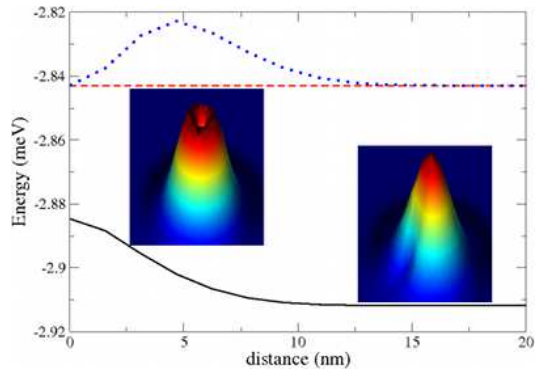}
\caption{\label{fig6} (color online) The lowest eigenenergies versus
  the distance of the impurity
  from the center of the vortex. The solid line is the ground state
  energy, the dashed line is the energy of the excited state whose
  wavefunction is shown in Fig. \ref{fig5} while the other excited
  state's wavefunction is shown in Fig.  \ref{fig5}
  (d). The left (right) inset is the ground-state wavefunction for
  $r_i=0$ ($r_i=5$ nm).}
\end{figure}

Next, we investigate the dependence of the bound energies and
wave-functions on the location of the impurity, for a fixed strength
$\alpha$. The results are shown in Fig.\ref{fig6}. The ground state
energy decreases monotonically as the impurity moves away from the
vortex center because it costs less energy to create a dip at its
location (the ground-state wavefunction is shown in the inset). Once the impurity
is farther away than the characteristic length scale of the bound
state, the ground-state energy saturates to its unperturbed value.
Due to the symmetry of the problem, the degeneracy of the first
excites state is lifted. The state which has its nodal line where the
impurity is located is unaffacted by it. The other eigenstate has a
finite wavefunction at the impurity location, and its energy is
increased. However, if the impurity is placed far enough, its effect
vanishes and the two excited eigenstates become degenerate again.  The
excited wavefunctions are shown in Figs.\ref{fig5}(b) and (d).

\begin{figure}
\includegraphics[width=1.0\columnwidth]{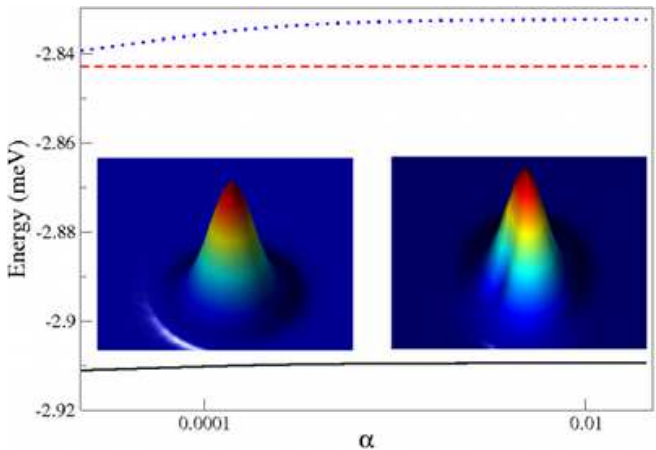}
\caption{\label{fig7} The ground, first and second excited state
  energy versus the impurity strength. The left inset is the
  wavefunction for $\alpha=0$, and the right inset is the wavefunction
  for $\alpha = 0.01$.}
\end{figure}

Finally, we analyze how the energies of the lowest eigenstates change
when the impurity is off-center at a fixed distance $r_i=5$ nm and we
vary the impurity strength.  The results are shown in
Fig. \ref{fig7}. As expected, the ground-state energy increases but
very little, since for large $\alpha$ a dip appears at the location
of the impurity and its effect saturates, irrespective of its
strength. The excited state with a nodal line at the impurity location
continues to be unaffected by it, while the other excited state
behaves similarly to the ground-state: its energy increases with
$\alpha$, but only by a finite amount until its wavefunction acquires a
zero where the impurity is.

These results clearly demonstrate that a single impurity is unable to
unbind the Zeeman trapped charge carrier, irrespective of how large
its repulsive potential is. All it can do is to simply ``poke a hole''
in the wavefunction, and this costs a relatively small energy. Even if
the impurity is placed at the center of the Zeeman potential well, the
energetic cost is little. This suggests that one could have many
impurities in the vortex area without affecting the bound state
significantly. Our expectations are indeed verified in the next
section.

\section{Random distribution of impurities\label{random}}

We now address a more realistic situation of a random distribution of
multiple impurities inside the QW.  In this case, as detailed in the
appendix, the use of B-splines is particularly useful to minimize
computational costs.

For a large number of impurities, the energy and wave function will
depend on the distribution of impurities. We follow the standard
procedure and average our results over various impurity
configurations. We performed averages over 1000 configurations at
different impurity concentrations.  The digitally doped
Cd$_{1-x}$Mn$_x$Te has a Zinc-Blende structure with a lattice constant
of about $6.5 \AA$. There are 4 Cd in an unit cell, and we take $x
\sim 1.6$. Assuming a single layer of Cd$_{1-x}$Mn$_x$Te gives a Mn
areal concentration of $0.15$ nm$^{-2}$. As already discussed, the
number of charged dopants can be minimized by the use of co-doping
techniques. While is impossible to eliminate all these scattering
centers, we expect that their concentration is much smaller than the
Mn concentration.

\begin{figure}[!ht]
\includegraphics[height = 1.9in]{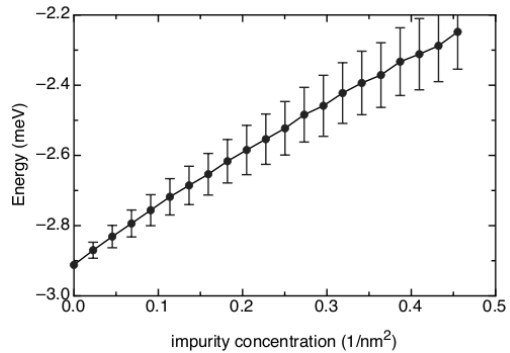}
\caption{\label{fig8} The ground state energy versus the impurity
  concentration. The energy was averaged over 1000 different impurity
  distributions and we chose $\alpha = 0.005$.}
\end{figure}

Fig. \ref{fig8} shows the central result of this work: the average
ground-state energy of the bound state increases linearly with the
concentration of impurities, however the slope is very small. Thus,
for any reasonable concentration of impurities we see that their
effect on the bound state energy is very minimal. This is fully expected,
given the analysis presented in the previous sections. Moreover, those
results also imply that increasing the relative strength $\alpha$ of
the scattering potentials will not change this conclusion, since the
effect of each scatterer saturates at large $\alpha$. 

Finally, we comment briefly on the effects of other types of
scatterers mentioned in the introduction. Clearly, attractive
impurities cannot unbind the spin-polarized charge carrier -- to the
contrary, they will bind it more strongly. Weak scatterers will have
small, perturbational effects on the bound state.  If the attractive
potential is extremely strong, a charge carrier becomes bound to the
impurity itself and will screen its potential, making the scatterer
``invisible'' to the spin-polarized charge carrier trapped by the
Zeeman potential. 
We conclude that such scatterers cannot damage the
bound state. 

The other potential source of scattering is due to ``noise'' in the
values of $g_{\rm eff}$ reflecting the variations in the local density of
Mn. One expects the typical scale for such variations (especially in
digitally doped layers) to be very short compared to the tens of nm
length scale of the bound wavefunction. If we model it in terms of a
sum of short-range (delta function) noise, the conclusions will be as
above, especially since the typical variation must be a small fraction
of the average $g_{\rm eff}$, i.e. we expect this to be in a
weak-scatterer regime.

\section{Conclusions\label{conclusions}}

In this article we studied the effect of single and multiple
impurities on the spin-polarized state bound by a trapping potential generated by an
inhomogeneous magnetic field in a diluted magnetic semiconductor QW. We
demonstrated that this effect is very limited, because of the large
difference in length scales between the size of the  wavefunction and
the typical size of a strong scattering potential. For very large
scattering potentials, each impurity pokes a hole in the bound
wavefunction, at a very small energetic cost. As a result, even for
 large concentrations of very strong scatterers, the overall effect on
 the bound state is very limited.

This result has obvious positive implications for any applications based on
such trapped states, since they show that it is not necessary to worry
about the effects of impurities or to use expensive methods to eliminate
them. Samples grown in reasonably clean conditions should suffice for
devices based on these spin-polarized, Zeeman-trapped states.

\acknowledgments T. G. R would like to acknowledge the Brazilian
agencies CNPq and FAPERJ and L'Or\'eal Brazil for the financial
support. 
B. J and S. L acknowledge the financial support of the
National Science Foundation NSF-DMR 06-201014, the US Department of
Energy Basic Energy Sciences and The Institute for Theoretical
Sciences, a Joint Institute of Argonne National Laboratory and the
University of Notre Dame. M. B. acknowledges support from the Research
Corporation, NSERC and CIfAR Nanoelectronics.

\appendix

\section{B-splines\label{appendix}}

B-splines have long been employed successfully  in atomic and molecular
physics,\cite{bsplines} but are not as freqently used in condensed matter
physics.\cite{splinessol} The aim of this appendix is to give a brief
intuitive picture 
of the B-splines and their usefulness for solving Schr\"odinger
equations with complicated potentials. 
Many more details on  B-splines and their uses are available in the
literature.\cite{bsplines}

B-splines are piecewise polynomials and therefore are well suited for
interpolation, having been extensively used in fitting tools,
including many commercial software applications.\cite{soft} In the
context of interest to us, they are used as a basis in which to expand
the eigenfunctions. Of course, there are many possible
finite basis sets and finite elements methods that can be used to
solve a Schr\"odinger equations.  The main advantage of using
B-splines is the flexibility to choose the grid points on which the
B-splines are defined. If we need to describe slowly-varying
functions, a large mesh suffices, resulting in a small basis set. In
regions where there are fast 
variations, one can use a finer local mesh, optimized to give the desired
accuracy for the minimum increase in the number of basis
functions. Furthermore, because the B-splines are piecewise polynomials,
matrix elements can be efficiently evaluated to machine accuracy with
Gaussian integration. Finally, the banded nature and sparsity of the
resulting matrices allows for the use of very large basis sets, if need
be.

Suppose that we need to approximate a 1D function in a given
interval $x \in [ a, b] $. We first define a knot-sequence of points
in this interval $\{x_i | a = x_0 \le x_1 \le x_2 \le ... \le x_N =
b\}$. The location of the points as well as their number will be
chosen so as to optimize the process. For instance, the knot-sequence
can consist of regions of  equally spaced knots combined with regions where
the knots are space in an exponential fashion, or any other suitable scheme. On
this knot-sequence, we define the normalized B-splines of rank $k$ by
the following recursion relation:
\begin{equation}
B_{i,1}(x) = \left\{
\begin{array}{c l}
1, & \mbox{if } x_i \le x < x_{i+1} \\ 0, & \mbox{otherwise}
\end{array}\right.
\end{equation}
and
\begin{equation}
B_{i,k}(x) = \frac{x - x_i}{x_{i+k-1}-x_i} B_{i,k-1}(x) +
\frac{x_{i+k} - x}{x_{i+k} - x_{i+1}} B_{i+1,k-1}(x)
\label{eqn:B}
\end{equation}
where $i$ is the index of a knot point and $k$ is the order of the
spline. Thus, the  $B_{i,k}(x)$ is a polynomial of order $k-1$ defined
piece-wise in the interval 
$[x_i,x_{i+k}]$, and vanishing outside it. Moreover, all derivatives
up to the order of $k-2$ are also continuous. Since we use 
these functions to expand eigenstates, which are continuous and have
first and second continuous derivatives, it follows we should use a
cubic ($k=4$) or higher order B-splines. Here we use cubic splines, as
they are the simplest splines with the desired properties. From now
on, we simplify our notation and use $B_i(x)$ to mean the cubic B-spline.
The function of interest is then expanded in terms of these splines:
\begin{equation}
f(x) = \sum_i \alpha_i B_i(x)
\end{equation}
where $\alpha$ are the $N+3$ coefficients of our expansion (for the
$N+1$-point knot sequence).

To solve an eigenvalue problem, we expand the wavefunction in terms of
B-splines and reduce the problem to a
general matrix system:
\begin{equation}
\mathbf{H}-\varepsilon\mathbf{S}\mathbf{v}=0,
\end{equation}
where $\mathbf{H}$ is the Hamiltonian matrix, $\mathbf{S}$ is the
overlap matrix and $\mathbf{v}$ is the eigenvector containing the
unknown coefficients for the various B-splines.  For a 1D
Schr\"odinger equation with a  potential $V(x)$, the matrix elements
of $\mathbf{H}$ and $\mathbf{S}$ are equal to:
\begin{eqnarray} 
H_{i,j}&=&\frac{1}{2m}\int { dB_i(x)\over dx}\frac{d B_j(x)}{dx}dx+\int
B_i(x)V(x)B_j(x) dx \nonumber \\ S_{i,j}&=&\int B_i(x)B_j(x) dx
\end{eqnarray}
where we used an integration by parts in the kinetic energy. The
integrals for the kinetic energy and the overlap can be evaluated
analytically, while those of the potential may need numerical
integration, if $V(x)$ is a complicated function. However, given the
finite support of each B-spline, such 
matrix elements are non-vanishing 
only if $|i-j|< k$, 
so the number of needed
integrals scales like the number of basis functions, not like its
square, as is the case for most other basis. 

\begin{figure}
\vskip 0.3 cm \includegraphics[height = 1.9in,clip]{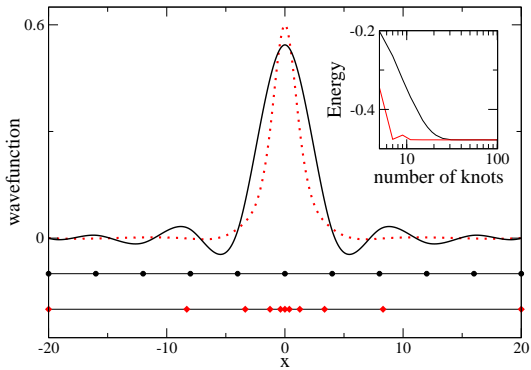}
\caption{\label{fig9} The dots on the  two lower lines illustrate the
uniform (squares) and non-uniform (diamonds)  knot sequences. The
curves show the estimated ground-state wavefunction for these $N=11$
knot sequences, full line for the uniform and the dotted line for the
non-uniform one. The inset shows the convergence of the energy to the
true value as a function of the number $N$ of knots used.}
\end{figure}

The choice of the knot sequence plays an important role in these
solutions. A good choice of knots distribution can assure a fast
convergence to the true eigenenergy with a small basis set. To
illustrate this, we consider a gaussian potential and we
calculate the ground-state
eigenvalue and eigenfunction using two different
knot sequences. As illustrated in Fig \ref{fig9}, one is 
a uniform distribution of knots, while the other has an exponential
distribution of knots from the center of the potential. In the inset
of Fig \ref{fig9}, we compare the convergence of the energy as a
function of the number of knots. It 
is clear that the non-uniform distribution is much more efficient in
this case, giving a very accurate wavefunction and eigenenergy for a
sequence with very few knots, {\em i.e.} a very small basis set. 

In the atomic physics, rigid boundaries are normally used when working
with B-splines. Since in solid state physics we sometimes want to
investigate bulk and transport properties, we opted to generalize the
B-spline method to periodic boundary conditions.  The periodic
conditions for B-spline are constructed in the following way (for 
cubic splines, although the generalization to higher orders is
straightforward). Suppose we want B-spines on the support of the knot
sequence 
$\{x_i | a = x_0 \le x_1 \le x_2 \le ... \le x_N = b\}$,
	
i) Expand the knot sequence $\{x_i\}$ to $\{x^{\prime}_i\}$ by adding
the knots:
\begin{eqnarray}
x^{\prime}_{-2} &=& x_{N-2} - (b-a)\\ x^{\prime}_{-1} &=& x_{N-1} -
(b-a).
\end{eqnarray}
Note that since $x^{\prime}_{-1}$ and $x^{\prime}_{-2}< a$, they are
outside the interval of interest.

ii) Construct the B-splines $\{B^0_i\}$ with the usual procedure on
the new support $\{x^{\prime}_i\}$. Note that the first two
B-splines, which we call $B^0_{-3}, B^0_{-2}$ have finite support outside our desired
interval.

iii) Now construct the B-splines $\{B_i\}$ by moving the 
pieces of $B^0_i$ defined outside the interval of interest, to inside $[a,b]$ by
a translation of period $(b-a)$. 

Following the above procedure, all $\{B_i(x)\}$ are periodic functions
with period $(b-a)$. Hence the functions in the Hilbert space expanded
by $\{B_i(x)\}$ with coefficients $c_i's$:
\begin{equation}
f(x) = \sum_i c_i B_i(x)
\end{equation}
are all periodic, thus ensuring the periodic boundary condition. Of
course, for our problem of interest we choose $b-a$ to be large enough
that the localized wavefunction is all fully contained inside it. In
other words, increasing $b-a$ has no effect on the eigenenergies we
calculate. 

The generalization to two-dimensional or higher-dimensional systems is
straightforward. The 
wavefunction needs to be expanded in products of B-splines
\begin{equation}
\psi(x,y)=\sum_i\alpha_{i,j}B_i(x)B_j(y)
\end{equation}
and all the procedures above can be repeated.  For our problem, we
simply choose a high-density of knots in the immediate neighborhood of
each impurity, where the wavefunctions change rapidly, and a wide,
exponential mesh everywhere else, where the function changes
slowly. The mesh near each scatterer has been optimized till
convergence is reached. 


\begin{references}
\bibitem{GaMnAs} H. Ohno, Appl. Phys. Lett. {\bf 69}, 262 (1996).

\bibitem{Furdyna1} J. K. Furdyna, J. Appl. Phys. {\bf 64}, R29 (1988).

\bibitem{Dietl1}T. Dietl, M. Sawicki, M. Dahl, D. Heiman,
  E. D. Isaacs, M. J. Graf, S. I. Gubarev, and D. L. Alov,
  Phys. Rev. B {\bf 43}, 3154 (1991).

\bibitem{03prl} M. Berciu, and B. Janko, Phys. Rev. Lett. {\bf 90},
246804 (2003).

\bibitem{Fiederling} R. Fiederling, M. Keim, G. Reuscher, W. Ossau
  G. Schmidt, A. Waag, and L. W. Molenkamp, Nature (London) {\bf 402},
  787(1999).

\bibitem{05apl} P. Redlinski, T. G. Rappoport, A. Libal,
  J. K. Furdyna, B. Janko, and T. Wojtowicz, Appl. Phys. Lett. {\bf
  86}, 113103(2005).

\bibitem{05prb} P. Redlinski, T. Wojtowicz, T. G. Rappoport, A. Libal,
  J. K. Furdyna, and B. Janko, Phys. Rev. B {\bf 72}, 085209(2005).

\bibitem{05nature} M. Berciu, T. G. Rappoport, and B. Janko, Nature
  (London) {\bf 435}, 71 (2005).

\bibitem{06prb} T. G. Rappoport, M. Berciu, and B. Jank{\'o},
  Phys. Rev. B {\bf 74}, 094502 (2006).

\bibitem{peeters} F. M. Peeters, and A. Matulis, Phys. Rev. B {\bf
  48}, 15166 (1993).

\bibitem{freire} J. A. K. Freire, A. Matulis, F. M. Peeters,
  V. N. Freire, and G. A. Farias, Phys. Rev. B {\bf 61}, 2895 (2000).

\bibitem{dms2deg}J. Jaroszynski et al., Phys. Rev. Lett {\bf 89},
  266802 (2002); B. Jussereand et al., Phys. Rev. Lett.  {\bf 91},
  086802 (2003); R. Knobel, N. Samarth, J. G. E. Harris and
  D. D. Awschalom, Phys. Rev.  B {\bf 65}, 235327 (2002); M. Goryca et
  al., Phys. Rev. Lett. {\bf 102}, 046408 (2009).

\bibitem{nanosup} J. Van de Vondel, C. C. de Souza Silva, B. Y. Zhu,
M. Morelle, and V. V. Moshchalkov, Phys. Rev. Lett. {\bf 94}, 057003
(2005)

\bibitem{marcel} C. Weeks, G. Rosenberg, B. Seradjeh and M. Franz,
  Nature Physics {\bf 3}, 796 (2007).

\bibitem{digitalDMS} R. W. Kawakami, E. Johnston-Halperin, L. F. Chen,
  M. Hanson, N. Guebels, J. S. Speck, A. C. Gossard, and
  D. D. Awschalom Appl. Phys. Lett. {\bf 77}, 2379 
  (2000); T. C. Kreutz, G. Zanelatto, E. G. Gwinn, and
  A. C. Gossard, Phys. Lett. {\bf 81}, 4766 
  (2002). 

\bibitem{bending} S. J. Bending, K. von Klitzing, and K. Ploog,
  Phys. Rev. Lett. {\bf 65}, 1060 (1990); A. K. Geim, S. J. Bending,
  and I. V. Grigorieva, Phys. Rev. Lett. {\bf 69}, 2252 (1992).
 
\bibitem{Kuhn1} B. Kuhn-Heinrich, W. Ossau, H. Heinke, F. Fischer,
T. Liz, A.  Waag, and G. Landwehr, Appl. Phys. Lett. {\bf 63}, 2932
(1993).

\bibitem{Lut}J. M. Luttinger and W. Kohn, Phys. Rev. {\bf 97}, 869
  (1955).

\bibitem{bsplines} H. Bachau, E. Cormier, P. Decleva , J. E. Hansen
  and F Martin, Rep. Prog. Phys. {\bf 64} (2001); J. Sapirstein and
  W. R. Johnson J. Phys. B: At. Mol. Opt. Phys. {\bf 29} (1996).
 
\bibitem{carneiro} G. Carneiro and E. H. Brandt, Phys. Rev. B {\bf
61}, 6370 (2000).
\bibitem{soft} See for example, Microcal Origin and Matlab.
\bibitem{splinessol} J. Sanchez-Dehesa, J. A. Porto, F. Agullo-Rueda
and F. Meseguer, J. Appi. Phys. {\bf 73} (1993).
\end{references}
\end{document}